\documentclass[twocolumn,10pt]{asme2ej}

\usepackage{epsfig} 
\usepackage{comment}
\usepackage[usenames,dvipsnames]{xcolor}
\usepackage{graphicx}
\usepackage{cancel}
\usepackage{amsmath}
\usepackage{amssymb}
\usepackage{bm}

\usepackage{pgfplots}
\usetikzlibrary{plotmarks}

\title{Going with the flow: enhancing stochastic switching rates in multi-gyre systems}

\author{Christoffer~R.~Heckman\thanks{Corresponding author}\\
    \affiliation{
U.S.\ Naval Research Laboratory\\
Code 6792\\
Plasma Physics Division\\
Nonlinear Dynamical Systems Section\\
Washington, DC 20375\\
    Email: christoffer.heckman.ctr@nrl.navy.mil
    }	
}

\author{M.~Ani~Hsieh\\
\affiliation{
Scalable Autonomous Systems Lab\\
Mechanical Engineering \& Mechanics\\
Drexel University\\
Philadelphia, PA 19104\\
        Email: mhsieh1@drexel.edu
    }
}

\author{Ira~B.~Schwartz\\
    \affiliation{U.S.\ Naval Research Laboratory\\
    Code 6792\\
    Plasma Physics Division\\
    Nonlinear Dynamical Systems Section\\
Washington, DC 20375\\
        Email address: ira.schwartz@nrl.navy.mil
    }
}

\begin{document}

\maketitle    

\begin{abstract}
{\it A control strategy is employed that modifies the stochastic escape
times from one basin of attraction to another in a model of a
double-gyre flow. The system studied captures the behavior of a large
class of fluid flows that circulate and have multiple almost
invariant sets. In the presence of noise, a particle in one gyre may
randomly switch to an adjacent gyre due to a rare large fluctuation.
We show that large fluctuation theory may be applied for controlling
autonomous agents in a stochastic environment, in fact
leveraging the stochasticity to the advantage of switching between
regions of interest and concluding that patterns may be broken or
held over time as the result of noise. We demonstrate that a
controller can effectively manipulate the probability of a large
fluctuation; this demonstrates the potential of optimal control
strategies that work in combination with the endemic stochastic
environment. To demonstrate this, stochastic simulations and
numerical continuation are employed to tie together experimental
findings with predictions.  }
\end{abstract}


\section{Introduction}
Time-dependent and stochastic environments like the ocean feature a 
significant downside for sensor operation: under the influence of the 
flow, the sensors will escape from their monitoring region of 
interest with some finite probability. To prolong operational 
life-spans, it is therefore desirable to design 
mobile sensing control strategies subject to minimizing the 
control effort.  For environmental monitoring applications, typically one
wishes to enable mobile sensors to either maintain their positions 
within a given region of interest or transition between regions to 
achieve more widespread sampling. For example, a common goal in a 
partitioned flow with multiple regions of interest is to either
discourage or encourage transitions between them
using an open-loop controller. In the former case, one strategy is to avoid
high probability regions of transition; in the latter, only actuating
an agent that happens into such a transition region will minimize effort
while optimizing the probability to transition between regions of interest
\cite{forgoston2011set,mallory2013distributed}. However, this approach
  generally can be inefficient;
oftentimes the controller does not take advantage of
subtleties of the flow field itself to optimize across both control
effort and transition probability.

Many systems are modeled with stochastic terms that represent uncertainties in 
model reduction \cite{doghre07,Forgoston_Chaos}, the variability in 
the time-dependent flow \cite{forgoston2011set}, and the fact that
position is sensitive to small fluctuations in the flow itself
\cite{vewaka08}. Included in such systems is a wide class of 
circulatory flows, with ``cells'' of rotating current that are 
produced by eddies or
vortices. These circulatory flows, commonly called gyres, are seen on
length scales as small as blood flow in the atria of a heart \cite{Fyrenius01} to
circulation between terrestrial features in the ocean  \cite{Anto}. Many
techniques exist to describe the qualitative aspects of such flows,
including identifying Lagrangian coherent structures
\cite{hall01,hall02,leshma07,inancacc05}, calculating finite-time Lyapunov exponents
(FTLEs) to pinpoint persistent but translating manifolds \cite{shlema05}, and
large-scale modeling of the fluid systems themselves.

To study the many systems which are modeled with stochastic effects, 
mathematical methods have been 
developed to elucidate the effects of noise on dynamical behavior
\cite{bbs02,bisc08,fd03,Froyland09,forgoston2011set}. These methods 
accurately predict the expected dwell 
time of agents within an individual region of interest before 
transitioning and the likely paths for
transitioning, including the most probable spatial exit points along the
basin boundaries. Stochastic methods may be used to classify those
phase space regions which are
almost invariant \cite{Froyland05} since deterministic basins of attraction
no longer exist. Such sets are regions in which agents in the
flow may remain for very long periods of time, but from which the agents
will eventually escape.

In contrast to set-based controllers, we will develop a control approach
that takes advantage of the flow governing the noise induced
escape from a region of interest. Specifically, we wish to know the actual trajectory which
has highest probability of escape. To design such a controller, we
will study the effect of small noise on the transition
times between two domains in a double-gyre flow. We will make use of 
the variational theory
of large fluctuations as it applies to finding the \emph{most
probable}, or \emph{optimal path} along which noise directs a particle to escape from an
almost invariant region \cite{dykman08}. This optimal path and its corresponding
switching rate will be used as a baseline in comparison with the switching rates
for multi-gyre flows with control.

It is well-known that noise has a significant effect on deterministic dynamical
systems. For example, consider a given initial state in the basin of
attraction for a given attractor, which might be steady, periodic, or
chaotic. Noise can cause the trajectory to cross the deterministic
basin boundary and move into another, distinct basin of
attraction~\cite{dyk90,dmsss92,mil96,lumcdy98}. For sufficiently small
noise, the dynamics are typically such that a particle first approaches
a stable manifold which defines the basin boundary, and then approaches a
saddle point on the basin boundary. Once the particle is near the
saddle, basin boundary crossings may occur randomly where noise
pushes the dynamics across the stable manifold after which point the
particle is carried along the unstable direction away from the
boundary. We note however that for large noise, such a crossing may
be determined by the global manifold structure away from the saddle
\cite{BillingsBS02}.

In the small noise limit, one can apply large fluctuation
theory \cite{feyhib65,dyk90,dmsss92,lumcdy98}, also known as large
deviation theory used in white noise analysis \cite{FW84,feyhib65}.
This approach enables one to determine the first passage times in a
vector field with multiple basins of attraction, and has
been applied to a variety of Hamiltonian and Lagrangian variational
problems \cite{wentzell76,Hu1987,Dykman1994d,FW84,GT84,MS93,HTG94}
outside of the context of control theory.

From a dynamical systems perspective, a subtlety to emphasize is that
systems under the influence of noise no longer have deterministic paths
that are governed by initial conditions. Instead, there is a probability
density function that describes the most probable position of hypothetical
particles within the flow. In stochastic systems that have multiple basins
of attraction, there exist ``most probable
paths'' for a hypothetical particle to transition between the two former
stable motions (whether they be equilibria, limit cycles, or other
dynamical behaviors). These transitional paths can be shown to be
exponentially more likely to occur than others \cite{feyhib65}, and as a result even in
systems that are under the influence of noise we can see through
theory and experiment that these paths are strongly preferred for
moving between adjoining basins. The presence of these paths enables 
us to exploit environmental forces to minimize control effort and 
therefore develop more energy efficient control strategies for small 
resource constrained mobile sensors operating in dynamic and 
uncertain environments. Large-fluctuation theory 
will enable us to identify these paths from a deterministic perspective
when the dynamics of the system are more relevant than the noise
intensity.

In this paper, we will show
that large fluctuation theory may be applied for controlling multiple autonomous
agents in a stochastic environment, in fact leveraging the stochasticity
to the advantage of switching between basins and concluding that
patterns may be held or broken over time as a result of noise 
\cite{MalloryNPG13}. The theory we develop will also be applicable to
noise sources that are non-Gaussian due to the general nature of the
formulation.

\section{The model}
Viewed from a kinematic perspective, a particle moving in a
fluid-driven multi-gyre flow may be modeled as \cite{YanLiu94}

\begin{equation}
\label{eom} \dot{\bm{q}} = \bm{u} + \bm{F}(\bm{q}) + \bm{\eta},
\end{equation}

\noindent where $\bm{q} = (x, y)$; $\bm{u}$ represents a control
response, and $\bm{\eta}$ is a stochastic white noise term with mean zero
and standard deviation $\sigma = \sqrt{2 D}$ for a given noise intensity $D$.
The delta-correlated noise is characterized by $\langle \eta_i(t) \rangle = 0$ and
$\langle \eta_i(t) \eta_j(t') \rangle = 2 D \delta_{ij} \delta(t-t')$.
The component of the gyre flow $\bm{F} = (F_1,F_2)$ represents the
deterministic, uncontrolled vector field given by

\begin{align}
        \label{gyre_flow1} F_1 &= -\pi A \sin(\pi x) \cos(\pi y/s) - \mu x\\
        \label{gyre_flow2} F_2 &= \pi A \cos(\pi x) \sin(\pi y/s) - \mu y.
\end{align}

\begin{figure}[ht!]
        \centering
        \includegraphics[width=3.5in]{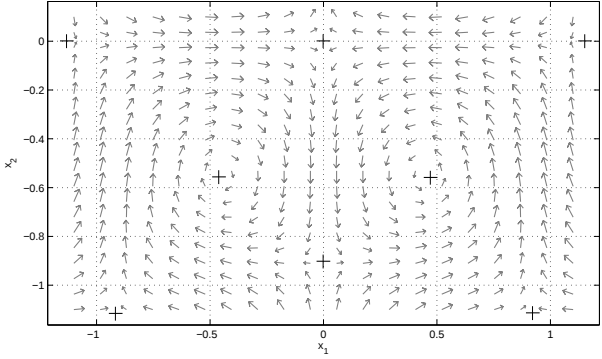}
        \caption{Phase portrait of the two-gyre vector field with
        arrows denoting the flow direction and positions of
        equilibria marked by $+$. The parameters for this simulated
        flow are $A = 1$, $s = 1$ and $\mu = 1$.}
        \label{fig:vectorfield}
\end{figure}

\noindent The parameter $\mu$ is a damping coefficient, $s$ is a
scaling dimension for the gyres, and $A$ corresponds to the strength
of the gyre flow. We note that undamped versions of the model were
studied in \cite{Rom-Kedar}, and that \cite{Froyland09} also examined
a forced, time-dependent version. A phase portrait of this system is
provided in Figure \ref{fig:vectorfield}.

The stochastic model will be used to quantify two situations in which noise
plays an important role: the first will be an examination of agent
escape times from the gyre without any control, and the second will
examine the effect of small gyroscopic controls to advance or retard
the escape times.
\subsection{The uncontrolled deterministic flow}

Consider the scenario without noise where $\bm{\eta} = 0$ and no controller is
present. In this case, the system is completely determined by
$\dot{\bm{q}} = \bm{F}(\bm{q})$, where $\bm{F}$ is given in Eqs.\
\eqref{gyre_flow1},\eqref{gyre_flow2}. The equilibria of
the deterministic, uncontrolled flow without damping (i.e.\ when $\mu
= 0$) are:

\begin{enumerate}
        \item the origin $\bm {q_0}=(0,0)$, which is exact.
        \item boundary equilibria: $(\pm 1, 0)$, $(0, \pm s)$, $(\pm 1,
                \pm s)$, $(\pm 1, \mp s)$.
        \item gyre equilibria: $(\pm 1/2, \pm s/2)$, $(\pm 1/2, \mp
                s/2)$.
\end{enumerate}

Note that in Eqs.\ \eqref{gyre_flow1}, \eqref{gyre_flow2},
substituting $x \rightarrow -x$, $y \rightarrow -y$ results in the
same equations of motion. Because of this symmetry, we will restrict
our investigation to two adjacent gyres. Without loss of generality, we
will consider switching between the two gyres which are centered at
$\bm{q}_{A_l}=(-\frac{1}{2}, -\frac{s}{2})$ and $\bm
{q}_{A_r}=(\frac{1}{2}, -\frac{s}{2})$, with the boundary point
through which the switching occurs being $\bm{q}_B = (0,-s)$.

When $\mu \neq 0$, the positions of the equilibria $\bm{q}_{A_l}$, 
$\bm{q}_{A_r}$, and $\bm{q}_B$ are perturbed.
Although we use numerical approximations to calculate the coordinates of
the equilibria, it is helpful to have an approximation so linearized
stability determined from eigenvalues may be determined. Applying
perturbation expansions for small $\mu$ to the gyre equilibria, their
positions to second order in $\mu$ are:

\begin{align}
        x^* &\approx \pm \frac{1}{2} \mp \frac{s}{2 \pi^2 A} \mu \mp
        \frac{s}{2 \pi^4 A^2} \mu^2\\
        y^* &\approx \pm \frac{s}{2} \pm \frac{s}{2 \pi^2 A} \mu \mp
        \frac{s^2}{2 \pi^4 A^2} \mu^2.
\end{align}

\noindent Using this approximation, the real part of these eigenvalues to second
order in $\mu$ is:

\begin{equation}
        \label{stability_gyre}
        \Re(\chi^*) = -\mu + (s-1)\frac{\mu^2}{8 A}.
\end{equation}

\noindent This suggests that, when they exist, the gyre equilibria 
are stable for $\mu > 0$.

Both the gyre and boundary equilibria are born out of saddle-node
bifurcations of the boundary equilibria and origin, respectively. Our
following analysis is only relevant for the parameter subset in
which all three types of equilibria exist and have these stability
properties.

For small $\mu > 0$, the origin always has one stable and one unstable
eigendirection. For the boundary equilibrium point at $(0,-s)$, we take the
linearization using the small $\mu$ approximation and find that the real parts
of the eigenvalues are:

\begin{equation}
        \label{boundary_eigs} \Re(\chi_1) = \pi^2 A - \mu - \frac{s^2
        \mu^2}{2 A} \qquad \Re(\chi_2) = -\frac{\pi^2 A}{s} - \mu +
        \frac{s \mu^2}{2 A}.
\end{equation}

\noindent This implies that there also will always be one stable and
one unstable eigendirection of this boundary equilibrium in the
parameter subset in which we are interested.

To summarize the phase portrait of a single gyre and its boundary, the
center of each gyre contains a stable focus and its basin of attraction is
bounded by four saddle equilibria. Adjacent to this gyre on all sides are
rotated and/or reflected replicas. Therefore, a system of two gyres sharing
a boundary qualitatively resembles a potential field with two potential
troughs separated by a peak. This illustration is helpful in considering
stochastic trajectories that can cause a particle to switch its position
from being in the basin of one gyre to that of another.

\section{Noise-induced gyre escape}
Under the influence of noise, the dynamical behavior of the system is
determined by its stationary probability density. In particular, all
equilibria are now peaks or troughs in a probability landscape
describing where a particle is likely to be located. There
are no orbits that are uniquely defined by a point along them, and
as a corollary invariant manifolds (such as the gyre boundaries) cease
to exist. Although there exist many paths that transport a particle
from one gyre to another in the presence of small noise, there are
``most likely paths'' in the sense that they will lie along a local
peak in the probability density. Such paths transitioning between
gyre basins will pass from a local maximum of the distribution
through a local minimum to a nearby local maximum. That is, the path
of transition will begin at an attracting state, escape from the
basin due to an effective force due to noise, and approach the other
attracting state in a different basin. We characterize this
transition by considering the most likely paths between the two states.

The probability of escape from an attractor under the influence of small
white noise scales exponentially as

\begin{equation}
        \label{noise_prob} \mathcal {P}_{\bm {\eta}}(\bm{q})=\exp(-\mathcal{R}(q))/D,
\end{equation}

\noindent where $\mathcal{R}$ is the action \cite{feyhib65}. Following the ideas in
\cite{FW84}, we notice that for any given realization of noise, we have
the approximate density scaling as $\exp(-\frac{1}{2}\int{|\bm \eta|^2})$. If
we continue the thinking of Feynman \cite{feyhib65}, then for any realization
of noise we get the action of white
noise defined in \cite{FW84}. We take a general Hamiltonian
approach which admits the formulation for escape induced by
non-Gaussian noise \cite{sbdl09}.

\subsection{Uncontrolled case}
To put the problem into Hamiltonian formulation, we characterize the
paths that require the minimum action of the dynamics and noise to
cause the transition. In this approach, we use the vector field of
Eq. \ref{eom}
as a constraint. We may then use calculus of variations to accomplish the
minimization. The action functional for the noise is:

\begin{equation}
        \mathcal{R}
        [\bm{q},\bm{\eta},\bm{\lambda}] =
        \frac{1}{2} \int \bm{\eta}(t) \cdot
        \bm{\eta}(t) dt + \int
        \bm{\lambda} \cdot (\dot{\bm{q}} -
        \bm{F}(\bm{q}) - \bm{\eta}) dt.\label{action}
\end{equation}

When evaluating the action in Eq. \ref{action}, we compute $R=\min\mathcal{R}
        [\bm{q},\bm{\eta},\bm{\lambda}]$, where the minimum is taken over the
        functions $[\bm{q},\bm{\eta},\bm{\lambda}]$.
Setting the first variation of the functional in Eq.\ \eqref{action}
to zero will yield a system of differential equations that identify
which solutions extremize the action in terms of the path and the
minimum noise necessary to realize the path. This solution is the
most probable path, even though it is rare and exists in the tail
of the probability distribution for realizations of the noise.

The switching rate to switch from one region to another is directly 
proportional to the probability of observing the most likely noise profile to induce
such a switch \cite{dykman08}; all other noise realizations are exponentially
less likely. Therefore, the switching time may be approximated as:

\begin{equation}
        \label{escape_scaling} T_S = b \exp\left(\frac{R}{D}\right)
\end{equation}

\noindent where $b$ is a prefactor which is determined through numerical
simulation or experiment and $R = \min \mathcal{R}$. In order to calculate the 
switching time in Eq.\ \eqref{escape_scaling}, we must find the optimal
trajectory and noise to induce a switch.

Computing the variational derivatives of the action and setting them
to zero results in the following differential equations:

\begin{align}
        \label{x_exp} \dot{x} &= F_1(x,y) + \lambda_1\\
        \label{y_exp} \dot{y} &= F_2(x,y) + \lambda_2\\
        \label{l1_exp} \dot{\lambda_1} &= -\frac{\partial F_1}{\partial x} \lambda_1
                - \frac{\partial F_2}{\partial x} \lambda_2\\
        \label{l2_exp} \dot{\lambda_2} &= -\frac{\partial F_1}{\partial y}
                \lambda_1 - \frac{\partial F_2}{\partial y} \lambda_2
\end{align}

\noindent where $\lambda_i$ represent the conjugate momenta to $x, y$.

The optimal path to transition between gyres is determined by the system above
combined with a set of boundary conditions describing the gyre and
boundary points. In particular the boundary conditions are 
$\bm{q}(t \rightarrow -\infty) = \bm{q}_{A_l}$ and $\bm{q}(t 
\rightarrow \infty) = \bm{q}_B$, with $\bm{\lambda}(t \pm \infty) = \bm{0}$.

Note that this is a deterministic system; its solution will dictate the
optimal path to transition between the two gyres.  Eqs.\
\eqref{x_exp}--\eqref{l2_exp} contain equilibria at all locations in
$(x,y)$ as the system in Eqs.\ \eqref{gyre_flow1}, \eqref{gyre_flow2}
with the conjugate momenta equal to zero. Linearizing about the gyre equilibria in Eq.\
\eqref{x_exp}--\eqref{l2_exp} and taking the real part of the
eigenvalues of the gives

\begin{align*}
        \Re(\chi_{1,2}) = -\mu + \frac{s-1}{8A} \mu^2\\
        \Re(\chi_{3,4}) = \mu - \frac{s-1}{8A} \mu^2.
\end{align*}

Note that both sets of eigenvalues share the same real part since they are complex conjugates;
also, the equilibrium point has saddle stability. As for the origin, we have

\begin{equation*}
        \Re(\chi_{1,2}) = - (\pi^2 A + \mu) \qquad
        \Re(\chi_{3,4}) = \frac{\pi^2 A}{s} - \mu.
\end{equation*}

The path connecting the equilibria is described as a two point boundary value
problem governed by Eqs.\
\eqref{x_exp}--\eqref{l2_exp} and the boundary conditions as explained
above. Given the fact that the gyre attractor, $\bm{q}_A$, and the
boundary saddle through which escape occurs, $\bm{q}_B$, are both saddle
  equilibria in the full set of equations of motion, finding the most optimal
  path mathematically requires identifying a heteroclinic orbit. To
  solve for the path numerically, we implement a numerical algorithm
  known as the IAMM \cite{lindley13}. We then employ
  numerical continuation using \textsc{Auto}'s HomCont \cite{auto07p}
to increase the accuracy of the approximation and to study the
behavior of the path using different parameter values.

It is important to note that the above exposition is analogous to the 
optimal trajectory generation problem where Eq.\ \eqref{action}  is 
the objective function and Eqs.\ \eqref{gyre_flow1}, 
\eqref{gyre_flow2} can be viewed as the vehicle kinematics. In this 
context, $x$ and $y$ are analogous to vehicle states and $\lambda_i$ 
are the control inputs. Different from the canonical optimal control 
problem \cite{bryson}, $\lambda_i$ represent the conjugate noise terms.

\subsection{Controlled case}
In the above analysis, we examined the eigenvalues of the equilibria
in the expanded phase space without control. Now, we will consider
the controlled system. The system equations are those in Eq.\
\eqref{eom}, including the gyre flow as given in Eqs.\
\eqref{gyre_flow1}, \eqref{gyre_flow2}, but now $\bm{u}$ represents a
gyroscopic control force. For brevity, define:

\begin{align*}
        f_1 &= -\pi A \sin(\pi x) \cos(\pi y/s)\\
        f_2 &= \pi A \cos(\pi x) \sin(\pi y/s).
\end{align*}

\noindent We synthesize a controller that modifies the optimal path and
therefore the transition probabilities. The controller we apply is:

\begin{equation}
        \label{gyro_control} \bm{u} = \bm{\omega} \times
        c \frac{[f_1, f_2, 0]}{||[f_1, f_2, 0]||},
\end{equation}

\noindent where $c$ is a control parameter and $[f_1, f_2, 0]$ represents
the conservative part of the vector field $\bm{F}$ with an augmented 
trivial third dimension. Following Mallory et al.\ 
\cite{mallory2013distributed}, we take the vector $\bm{\omega}
= [0, 0, 1]^T$; i.e.\ it is a unit vector pointing into or out of the
plane of the flow depending on the flow direction. The definition of $\bm{u}$
imposes a controlled modification of the two-dimensional flow field.

Carrying out the cross product in Eq.\ \eqref{gyro_control}
and plugging the final result into the vector field, we have:

\begin{align}
        \label{control_flow1} \dot{x} &= -c
        \frac{f_2}{\sqrt{f_1^2 + f_2^2}} + f_1 - \mu x + \eta_1\\
        \label{control_flow2} \dot{y} &= c
        \frac{f_1}{\sqrt{f_1^2 + f_2^2}} + f_2 - \mu y + \eta_2.
\end{align}

Recall that the action is logarithmically related to the probability
of occurrence of escape. Since we are operating in the limit of small
$D$, it guarantees that for a sufficiently small noise intensity,
switching will only occur as a rare event and if switching should
occur, it will be as a result of this particular event.

To compare switching rates in the controlled versus uncontrolled case, we choose two
values of $c$---one positive and one negative---as examples for regimes in which
the particle is meant to be ejected or contained from the present gyre respectively. The
two values of the control constant and their corresponding values of the action are
given in Table \ref{table:controller_vs_action}.

\begin{table}
        \caption{Comparison of calculated values of $\mathcal{R}$ for two values of the
        control constant and the baseline case without control. The two nonzero values
        of $c$ were arbitrarily chosen from a set of paths computed using continuation.}
        \centering
\begin{tabular}{| c | c | c |}
        \hline
        $c$ & $\mathcal{R}$ \\ \hline
        $ 0.0967 $ & 0.0977 \\
        $ 0 $ (no control) & 0.161 \\
        $-0.314 $ & 0.379 \\
         \hline
\end{tabular}
\label{table:controller_vs_action}
\end{table}

Note that as $c$ is increased, the action $\mathcal{R}$ decreases.
Therefore, the particle is more likely to escape from one gyre to
another with the controller turned on.

\section{Simulation Results}

We consider two methods to compare the theory developed in the previous
section with numerical experiments: direct comparison of probability
density functions of numerically simulated
paths that lead to switching with the theoretically predicted paths, and comparison of the mean first passage
time (MFPT) from one gyre to another via simulation and that obtained
using the predicted estimate in Eq.\ \eqref{escape_scaling}.

Numerical simulations were obtained via stochastic integration using
an explicit Milstein method \cite{gar03}. All simulations started with initial
conditions at the center of the gyre $\bm{q}_A$, and were run until
they crossed the line $x = 0$, at which point the particle's
trajectory and the total time of the integration were recorded. Using
these two measures, we are able to compare the anticipated MFPT
and the path itself with expected results from large fluctuation
theory. All stochastic simulations were run in batches of 3,000 trials.

\begin{figure}[ht!]
        \centering
        \includegraphics[width=3.5in]{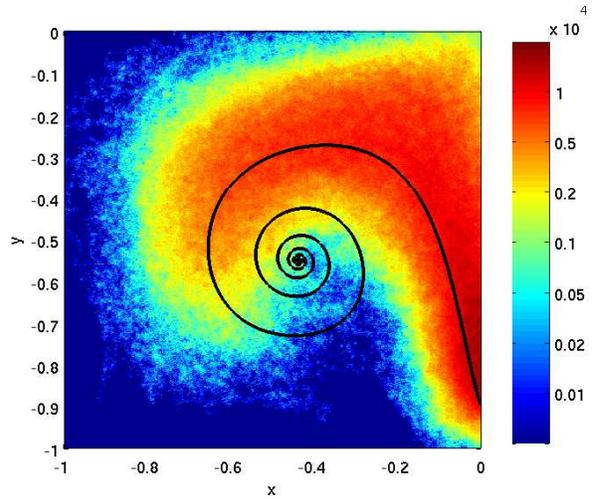}
        \caption{Optimal switching path for $c = 0$ overlaid on
        a probability density function (pdf) of paths that have been 
        stochastically integrated until
        transitioning out of the basin of attraction. The colormap represents
        an exponential scale of 3,000 sample paths with $D = 1/30$}
        \label{fig:nocontrol}
\end{figure}

\begin{figure}[ht!]
        \centering
        \includegraphics[width=3.5in]{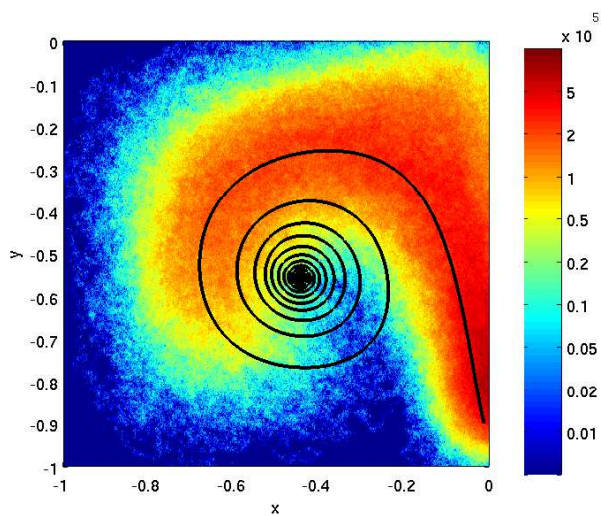}
        \caption{Optimal switching path for $c \approx 0.1$ overlaid on
        a pdf of paths that have been stochastically integrated until
        transitioning out of the basin of attraction. Paths were
        generated by the same method as in Fig.\ \ref{fig:nocontrol}}
        \label{fig:positive_c}
\end{figure}

In Figure \ref{fig:nocontrol}, the optimal path to escape is shown for
$c=0$, whereas in Figure \ref{fig:positive_c}, the path is
shown for $c = 0.0967$. The two-dimensional histograms over which these paths are
imposed shows the stochastic simulations of paths that have passed to
the adjacent gyre. Note that red indicates a higher incidence of occurrence and
blue shows less occurrence. The full histories of these paths have been truncated
and the results are shown on a logarithmic scale to accentuate
the ridge of higher probability.

\begin{figure}[ht!]
        \centering
        \includegraphics[width=3.5in]{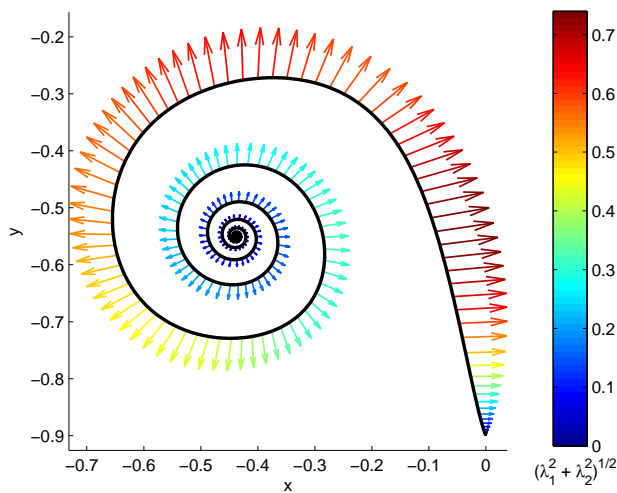}
        \caption{The conjugate momenta along the optimal path as generated
        by \textsc{Auto} for $c = 0$. The arrows indicate the direction in
        which the optimal noise is acting at the given point along the
        optimal path, and the color and length of the arrows indicate the
        magnitude of the noise.}
        \label{fig:conjugate_momenta}
\end{figure}

Figure \ref{fig:conjugate_momenta} shows a plot of the
conjugate momenta versus time along the optimal path as computed using
numerical continuation. This profile of the noise shows that it is optimally
always acting to push the particle outward, strengthening as the distance from the
center increases until it reaches the saddle point.

\begin{figure}[ht!]
        \centering
        \includegraphics[width=3.5in]{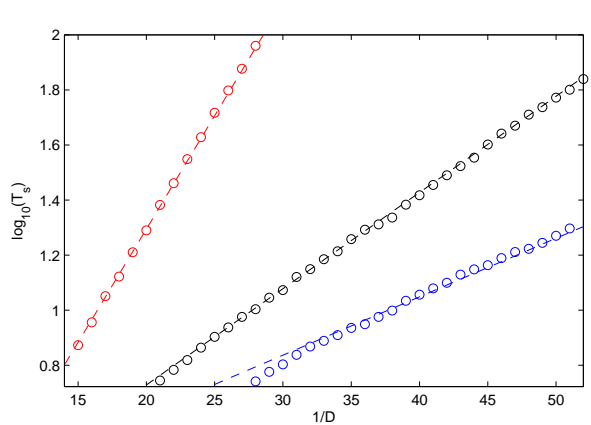}
        \caption{Comparison of the MFPT as predicted by large-fluctuation theory
        (dashed lines) and computed from an average over many
        stochastic trials (points). Three regimes were examined: black
        represents no control, red represents $c \approx -0.31$ and blue represents
        $c \approx 0.1$}
        \label{fig:switching_times}
\end{figure}

Finally, Figure \ref{fig:switching_times} compares the MFPT of
a particle via stochastic simulation and that expected using the theory.

\section{Conclusions and Discussion}
In this paper we have discussed a new approach to analyzing the
control of systems under the influence of weak noise by utilizing a
controller that manipulates the residence time of an agent
switching between adjacent basins of attraction. Because of the
stochasticity inherent to the dynamics of the flow, there will be a
rare but non-negligible chance that an agent will inadvertently
drift out of the region in which it should remain; this may be aided
or abated by a controller that aims to maximize or minimize the mean
first passage time between the basins. Both contexts are imperative
in many cases. In the example where there are many inexpensive,
independently-acting vessels that are tasked to uniformly cover a
domain composed of multiple basins of attraction, random
redistribution of the vessels will occur via stochastic effects. It
is advantageous in this example to choose a vessel that would be
easier to propel into an adjacent cell, apply a weak controller that
works in conjunction with noise to facilitate a transition, and on
all the other vessels apply control so as to impede a transition. We
have shown that with a simple, gyroscopic control approach this is a
feasible approach.

To show that efficacy of this approach, we analyzed the mean first
passage time from one basin to an adjacent one using a gyroscopic
control strategy. Specifically, we showed there exists a logarithmic
scaling of the mean escape times from one basin to another as a function
of the noise in the environment. Such scalings persist without control,
and the MFPT may be increased or decreased depending on the control
settings. Even though the controller never changed the stability of
the equilibrium point, there was a noticeable effect on the
transition rate as witnessed in the stochastic simulations. We
also showed that the logarithm of the transition rates need not be
calculated using
massive stochastic simulations, but rather can be approximated by
using large fluctuation theory. This enables us to predict mean
first-passage times and obtain an approximation for the reliability
of the controller in a stochastic environment.

In deploying a fleet of many small minimally propelled vessels, one 
objective is to establish a given (possibly non-uniform) spatial 
distribution of buoys throughout the domain of interest. If the 
domain is similar to multi-gyre flows, then it may be necessary to 
make the buoys switch from one gyre to another if the density is too 
high in a particular region. Simulations were run for negative and 
positive $c$ values in Eq.\ \eqref{gyro_control} in order to 
demonstrate that the controller can either maintain or change its 
gyre position. Since switching will always occur (albeit rarely) with 
no control imposed, it is advantageous to decrease the likelihood of 
a random switch---and therefore applying the control with $c<0$ will 
help maintain the status quo.

The results from both numerical simulations and large fluctuation
theory confirm this hypothesis. Both approaches show that the
MFPT for a particle is decreased for $c>0$ and is increased for $c<0$.
That is, when the control is focused inward then the particle
resides inside its original gyre for much longer than if the control is
focused outward. Since in a truly stochastic environment there will
always be unwanted transitions between basins of attraction, gyroscopic
control has been shown to manipulate the transition rate so a
targeted spatial distribution may be attained \cite{mallory2013distributed}.

We will also note that in reference to Figures \ref{fig:nocontrol}
and \ref{fig:positive_c}, while the paths near the gyre are difficult to
resolve, the transition paths near the boundary point is in agreement
with the optimal path predictions. By decreasing $D$ we can more
effectively resolve the sample histories and show agreement with
the large-fluctuation prediction, but the computation time increases
exponentially for small decreases in the noise intensity. The results
we show display a
trade-off between these factors, but it still captures the most salient
feature of the analysis: there is a strong preference to approach the
boundary point along a particular trajectory. This fact may be leveraged
in the design of other controllers that might aim to expend effort to
following along this trajectory and instigate a boundary crossing
with minimal energy expenditure.

Another noteworthy result is that the optimal paths for $c > 0$ are more
tightly wound than for $c = 0$; i.e.\ the path that contributes to the
larger action in fact requires a very large fluctuation in order to
cross into the adjacent basin. While at first this may be surprising, it
may be explained by the qualitative action of the controller. Since for $c>0$ the
controller encourages leaving the center of the gyre, a series of
more-likely smaller fluctuations (rather than a single very large one) will enable
a particle to overcome the effective potential barrier and escape the
basin.

These results have important implications for the design of
controllers in the presence of noise. A vector field when influenced by
stochastic effects retains many of the same dynamical behaviors of the
original system, but there is the possibility for random switching
events. With the knowledge of which paths are most likely to lead to
a transition, we may pick our controller constants in order to minimize
the MFPT. This is particularly important because it permits the efficient 
use of fuel thus allowing for much longer deployment times for each vessel.

In the future, we will consider the design of controllers unique to
the noise characteristics of the system as well as the underlying
vector field. One of the results in this study was the identification
of a most-likely path to switch between basins of attraction; this
path is of critical importance as shown in the simulation-generated pdf of
transitioning trajectories. While the controller we employed did not
explicitly make use of this path, we aim to replace it with one that
does---in particular, exerting significant control effort when it is
clear that a transition is likely to occur, but remaining nearly
dormant when a transition is unlikely.

\section{Acknowledgements}

This research was performed while
CRH held a National Research Council Research Associateship Award
at the U.S.~Naval Research Laboratory. This research is funded by the 
Office of Naval Research contract F1ATA01098G001, Naval Research Base 
Program Contract N0001412WX30002, and NSF grant IIS-1253917.

%
% ...

\bibliographystyle{asmems4}



\end{document}